\documentclass[12pt,a4paper]{elsart}
\usepackage{latexsym,amssymb,epsfig}
\usepackage{amsmath,eepic}
\newcommand{\di}{\ensuremath{\mathrm{d}}} 

\newcommand{\0}{$\phantom{0}$}

\def\0{\phantom{0}}
\renewcommand{\thefootnote}{\fnsymbol{footnote}}
\renewcommand{\thetable}{\Roman{table}} 
\begin{document}
\begin{center}
{\bf \large Transport Properties of Anisotropic Polar Fluids:}

{\bf \large 1. Quadrupolar Interaction}
\bigskip

G.A. Fern\'{a}ndez, J. Vrabec\footnote{To whom correspondence should be
addressed, tel.: +49-711/685-66107, fax: +49-711/685-66140, 
email: vrabec@itt.uni-stuttgart.de}, and H. Hasse

Institute of Thermodynamics and Thermal Process Engineering,

University of Stuttgart, D-70550 Stuttgart, Germany

\end{center}

\renewcommand{\thefootnote}{\alph{footnote}}
\baselineskip25pt

\vskip3cm Number of pages: 45

Number of tables: 1

Number of figures: 14

\clearpage
{\bf ABSTRACT}\\

Equilibrium molecular dynamics simulation and the Green-Kubo
formalism were used to calculate self-diffusion coefficient, shear viscosity,
and thermal conductivity for 30 different quadrupolar two-center Lennard-Jones
fluids along the bubble line and in the homogeneous liquid. It was systematically
investigated how anisotropy, i.e. elongation, and quadrupole momentum influence
the transport properties. 
The reduced elongation $L^*$ was varied from 0 to 0.8 and the reduced squared
quadrupole momentum $Q^{*2}$ from 0 to 4, i.e. in the entire range in which
parameters for real fluids are expected. 
The statistical uncertainty of the reported data
varies with transport property, for self-diffusion coefficient data 
the error bars are typically lower than $3 \%$, for shear
viscosity and thermal conductivity they are about 8 and $12 \%$, respectively.


\bigskip

\clearpage
\section{INTRODUCTION}
From the pioneering work of Alder et al. \cite{alder1,alder2} who used
molecular dynamics to investigate transport properties of hard-spheres, this
method has proved to be a successful tool for the development and test of theories
\cite{hansen}; examples are the discovery of the hydrodynamic long-time tail
\cite{alder3} or the development of advanced kinetic theories
\cite{furtado,keyes}. Due to the increasing computer power, molecular dynamics is today
an interesting option for studying \cite{macdowell} and predicting \cite{McCabe}
transport properties. In particular, for predictions of transport properties of
liquids, where no satisfactory analytical theory exists, molecular dynamics
is the most suitable available method.

Molecular simulation is also attractive because it allows rigorous testing of
theories and models, and systematically studying the influence of any molecular
parameter on any transport property. Although real fluids usually consist of polar
non-spherical molecules, most extensive studies on transport properties have been
done on the basis of very simple molecular models, e.g. hard sphere potential
\cite{alder2,erpenbeck1}, square well potential
\cite{michels1,michels2,michels3,michels4,michels5,michels6}, steeply repulsive
potential \cite{heyes1}, or spherical Lennard-Jones potential \cite{heyes1,heyes2}.

In the present work, a systematic study on the influence of elongation and quadrupole
momentum on self-diffusion coefficient, shear viscosity, and thermal conductivity of
two-center Lennard-Jones plus point quadrupole (2CLJQ) fluids is carried out in the liquid
region covering a broad range of temperature and density. This work is based on the knowledge of accurate
vapor-liquid equilibria of 2CLJQ fluids from previous publications of Stoll et al.
\cite{stoll1,stoll2}. Molecular simulations are carried out along the bubble line 
and in the homogeneous liquid for different
models over a grid of reduced temperatures and densities. 
Thus, a consistent comparison of the transport properties of different models is possible 
and subsequently the effect of elongation and quadrupole momentum can be identified.

It has been recently shown that the 2CLJQ model is not only an interesting model fluid but also suited for
accurately describing the properties of real fluids. Vapor-liquid equilibria of 35 pure substances were
successfully modeled with that approach \cite{vrabec,stolljcp}. Also for mixtures good results were
achieved \cite{stoll2,vrabec2,stolla}. Furthermore, properties such as Joule-Thomson inversion \cite{vrabec3},
self-diffusion and binary Maxwell-Stefan diffusion coefficients \cite{fernandez2}, shear viscosity, and
thermal conductivity \cite{fernandez3} were reliably predicted by 2CLJQ models which were parameterized
using vapor-liquid equilibrium data only.

\section{MOLECULAR MODEL}
The intermolecular interactions are represented by the two-center Lennard-Jones plus point quadrupole (2CLJQ) potential.
The 2CLJQ potential is pairwise additive and consists out of two identical Lennard-Jones sites a
distance $L$ apart (2CLJ) plus a point quadrupole of momentum $Q$ placed in the geometric center of the
molecule oriented along the molecular axis connecting the two Lennard-Jones (LJ) sites. The interaction
energy of two molecules $i$ and $j$ is given by

\begin{equation}\label{tpoten}
u^{\rm 2CLJQ}_{ij} = \sum_{a=1}^{2} \sum_{b=1}^{2} 4\epsilon \left[ \left( \frac{\sigma}{r_{ab}}
\right)^{12} - \left( \frac{\sigma}{r_{ab}} \right)^6 \right] + u_{\rm Q}.
\end{equation}

Here, $r_{ab}$ is one of the four LJ site-site distances; $a$ counts the two LJ sites of molecule
$i$, $b$ counts those of molecule $j$. The LJ parameters $\sigma$ and $\epsilon$ represent size
and energy, respectively. The contribution of a point quadrupole is given by \cite{graygubb}

\begin{eqnarray}\label{qpoten}
u_{\rm Q}=\frac{1}{4 \pi \epsilon_0}\frac{3}{4} \frac{Q^2}{\left|\textbf{r}_{ij}\right|^5}
\left[ 1-5 \left( c_i^2+c_j^2 \right) -15 c_i^2 c_j^2 + 2 \left( s_i s_j c - 4 c_i c_j
\right)^2 \right],
\end{eqnarray}

with $c_k={\rm cos} \theta_k$, $s_k={\rm sin} \theta_k$, and $c={\rm cos}
\phi_{ij}$. Herein, $\textbf{r}_{ij}$ is the center-center distance vector of two
molecules $i$ and $j$. $\theta_i$ is the angle between the axis of the molecule
$i$ and the center-center connection line and $\phi_{ij}$ is the azimuthal angle
between the axis of molecules $i$ and $j$. Finally, $\epsilon_0$ is the electric
constant in 8.854187817$\cdot$10$^{-12}$ C$^2$/(J m).

A specific 2CLJQ model, e.g. for a real fluid like nitrogen, is fully
determined by five parameters: $\sigma$, $\epsilon$, $L$, $Q$ \cite{vrabec,stolljcp} and the
molecular mass $m$. But in molecular simulation all relevant physical
properties can be treated in a reduced form. Here, they are related to $\sigma$, $\epsilon$,
and $m$, so that the reduced results are valid for all combinations of these three
parameters. In this form, only two molecular parameters remain, i.e. reduced elongation
$L^*=L/\sigma$ and reduced squared quadrupole momentum
$Q^{*2}=Q^{2}/(4\pi\epsilon_0\epsilon\sigma^5)$.
Henceforth, "squared" will be omitted in the text for brevity.

\section{TRANSPORT COEFFICIENTS}

Transport properties were calculated by equilibrium molecular dynamics simulation and the Green-Kubo
formalism \cite{green,kubo}. In this formalism, transport coefficients are obtained by integrating time
autocorrelation functions of the corresponding microscopic fluxes.

The self-diffusion coefficient $D$ is a measure for the mobility of individual molecules within a fluid.
It is calculated by integration of the single molecule velocity autocorrelation function
\cite{gubbins1,steele}

\begin{equation}\label{self}
D=\frac{1}{3N}\int_0^{\infty} \di t~\big\langle \mathbf{v}_k(t)\cdot \mathbf{v}_k(0) \big\rangle,
\end{equation}

where $\mathbf{v}_k(t)$ expresses the velocity vector of molecule $k$ at some time $t$, and $<$...$>$
denotes the ensemble average. Eq. (\ref{self}) yields the self-diffusion coefficient averaging over $N$ molecules.

The shear viscosity $\eta$, as defined in Newton's "law" of viscosity, describes the resistance of a fluid
to shear forces. It refers to the resistance of an infinitesimal volume element to shear at constant
volume \cite{baron}. The shear viscosity can also be related to momentum transport under the influence of
velocity gradients. From a microscopic point of view, the shear viscosity can be calculated by integration
of the time-autocorrelation function of the off-diagonal elements of the stress tensor, i.e. $J_p^{xy}$
\cite{gubbins1,steele}
\smallskip
\begin{equation}\label{shear}
\eta=\frac{1}{Vk_BT}\int_{0}^{\infty} \di t~\big\langle J^{xy}_p(t)\cdot J^{xy}_p(0)\big\rangle,
\end{equation}
where $V$ is the molar volume, $k_B$ is the Boltzmann constant, $T$ the temperature, and $<...>$ denotes
the ensemble average. The statistics of the ensemble average in Eq. (\ref{shear}) can be improved using
all three independent off-diagonal elements of the stress tensor, i.e. $J_p^{xy}$, $J_p^{xz}$, and
$J_p^{yz}$. For a pure fluid, the component $J_p^{xy}$ of the microscopic stress tensor $\textbf{J}_p$ is
given by \cite{evans1}
\bigskip
\begin{equation}\label{shear_t}
J_p^{xy}=\sum_{i=1}^N m v_{i}^x v_{i}^y -\frac{1}{2}\sum_{i=1} ^N \sum_{j \neq i}^N \sum_{k=1}^3
\sum_{l=1}^3 r_{i j}^x \frac{\partial u_{ij}}{\partial r_{k l}^y}.
\end{equation}
Here, $i$ and $j$ are the molecular indices. Lower indices $l$ and $k$ count all sites, including the
quadrupolar site, and the upper indices $x$ and $y$ denote the vector component, e.g. for velocity $v_i^x$
or center-center distance $r_{ij}^x$.

The thermal conductivity $\lambda$, as defined in Fourier's "law" of heat conduction, characterizes the
capability of a substance for molecular transport of energy driven by temperature gradients. It can be
calculated by integration of the time-autocorrelation function of the elements of the microscopic heat
flow $J^x_q$ and is given by \cite{gubbins1,steele}
\bigskip
\begin{equation}\label{termalc}
\lambda=\frac{1}{Vk_BT^2} \int_{0}^{\infty} \di t~\big\langle J^x_q(t)\cdot J^x_q(0) \big\rangle.
\end{equation}

The expression for the heat flow $\textbf{J}_q$ in pure fluids has been derived by Evans \cite{evans1} and
is given by
\bigskip
\begin{multline}
\label{qflux} \textbf{J}_q=\frac{1}{2} \sum_{i=1}^N \big((m v_i^2+ \textbf{w}_i \textbf{I}_i
\textbf{w}_i+\sum_{j \neq i}^N u_{ij} )\cdot \textbf{v}_i \big) -\frac{1}{2}\sum_{i=1}^N \sum_{j \neq i}^N
\sum_{k=1}^{3} \sum_{l=1}^{3} \textbf{r}_{ij} \cdot \big( \textbf{v}_i \frac{\partial u_{ij}}{\partial
\textbf{r}_{kl}}+\textbf{w}_i \mathbf{\Gamma}_{ij} \big),
\end{multline}

where $\textbf{w}_i$ is the angular velocity vector of molecule $i$, $\textbf{I}_i$ its matrix of angular
momentum of inertia, and $u_{ij}$ the intermolecular potential energy. The torque $\mathbf{\Gamma}_{ij}$
refers to a reference frame with origin in the molecular center of mass.  As for shear viscosity, all
three independent heat flow directions $J_q^x$, $J_q^y$, and $J_q^z$, can be used to improve the
statistics of Eq. (\ref{termalc}).

All results of this study were obtained and are presented in the reduced form, i.e. in relation to the
molecular parameters size, energy, and mass. The reduced transport properties are defined by:
$D^*=D/\sigma\sqrt{m/\epsilon}$, $\eta^*=\eta\sigma^2/\sqrt{m\epsilon}$,
$\lambda^*=\lambda\sigma^2 /k_B\sqrt{m/\epsilon}$. Relevant static thermodynamic properties, temperature
$T^*=Tk_B/\epsilon$ and number density $\rho^*=\rho\sigma^3$, are also reduced in the same sense. For the sake of
brevity, "reduced" will be omitted in the following.

\section{INVESTIGATED MODELS AND STATES}
\begin{figure}[ht]
\caption[Phase diagrams for two selected elongated 2CLJQ fluids ($L^*$=0.2) where one is non-polar and the
other strongly quadrupolar. Saturated densities, taken from \cite{stoll2}, are represented by the lines
joining at the critical point depicted by $\bullet$. The investigated state 
points are indicated by
$\circ$ for $Q^{*2}$=0 and by $\bigtriangleup$ for $Q^{*2}$=4.]{} \label{fig1}
\begin{center}
\includegraphics[width=150mm,height=200mm]{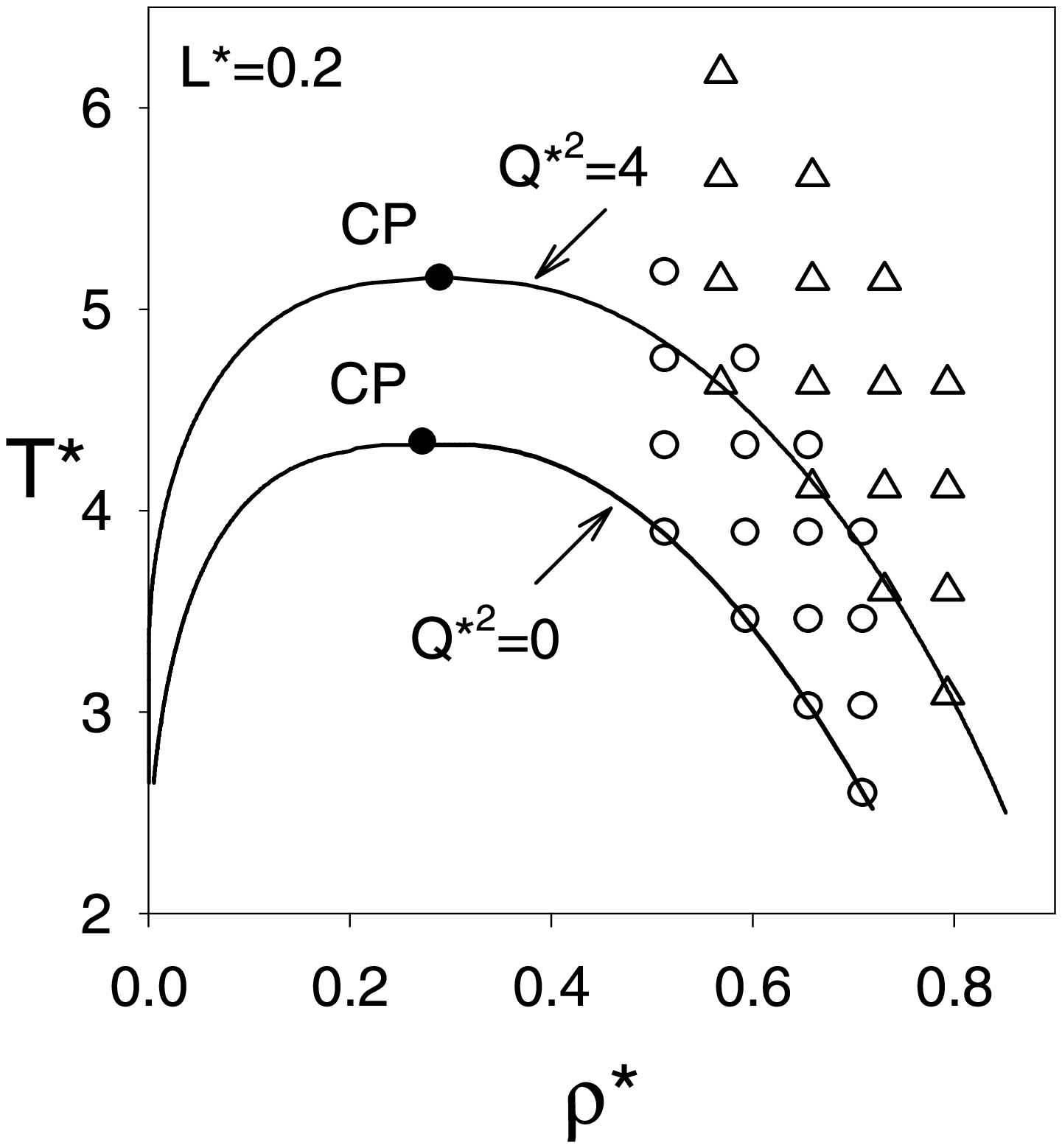}
\end{center}
\end{figure}

\begin{figure}[ht]
\caption[Phase diagrams for two selected quadrupolar 2CLJQ fluids ($Q^{*2}$=1) where one is spherical and
the other strongly elongated. Saturated densities, taken from \cite{stoll2}, are represented by the lines
joining at the critical point indicated by $\bullet$. The investigated state points 
are indicated by
$\circ$ for $L^*$=0 and by $\bigtriangleup$ for $L^*$=0.8.]{} \label{fig2}
\begin{center}
\includegraphics[width=150mm,height=200mm]{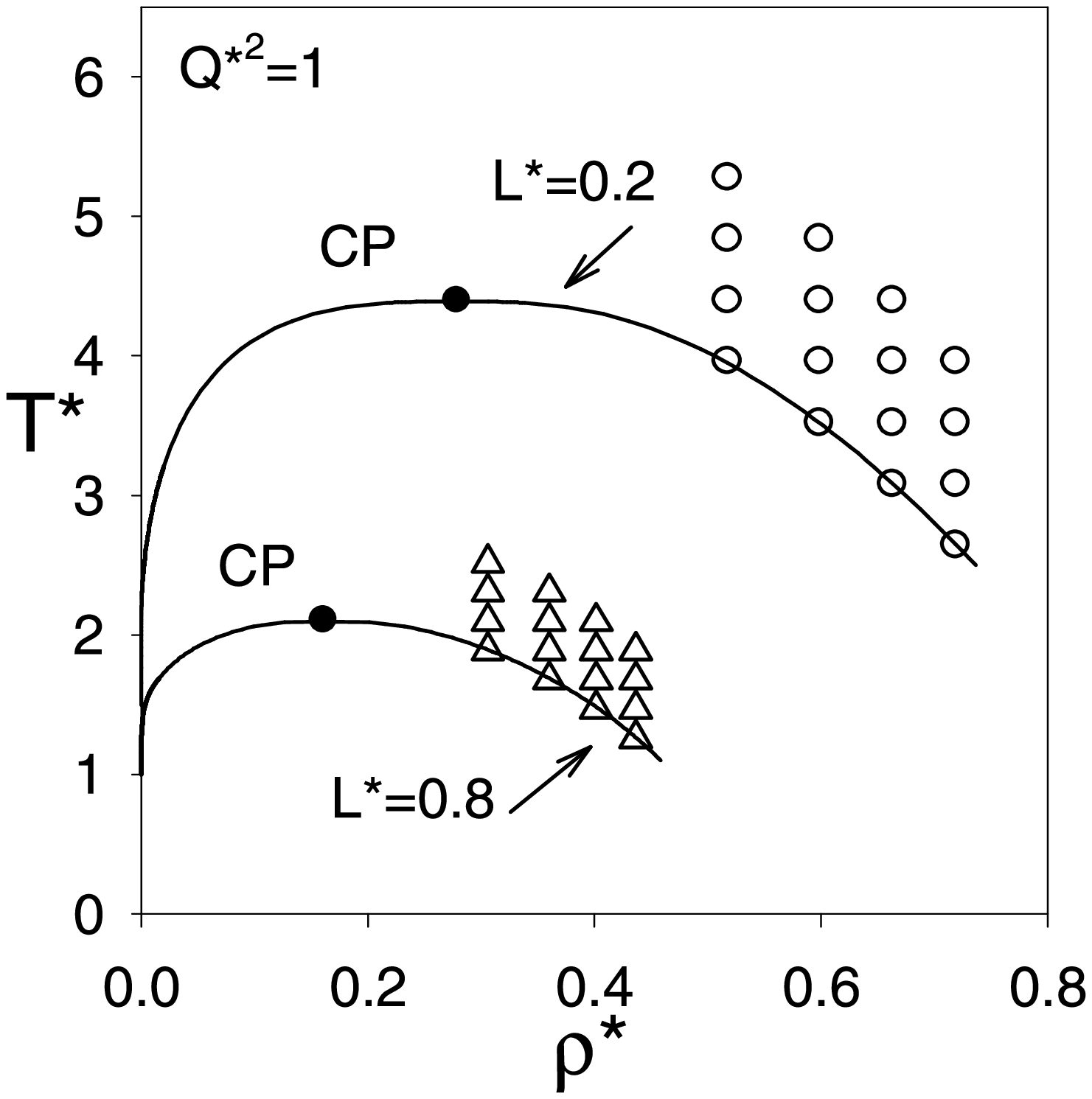}
\end{center}
\end{figure}

In the present work, 30 different model fluids were studied, where each fluid is fully
determined by one combination of elongation $L^*$ and quadrupole momentum $Q^{*2}$. 
Simulations at 16 liquid state points were carried out for each model fluid.
In Figs. \ref{fig1} and \ref{fig2} four selected systems are shown; they
illustrate the covered thermodynamic states and the influence of elongation and
quadrupole momentum on the thermodynamic behavior of the fluids \cite{stoll1}. As Fig.
\ref{fig1} shows, critical temperature and density increase with increasing
quadrupole momentum. On the other hand, Fig. \ref{fig2} shows that these critical
properties decrease with increasing elongation. Such a behavior can also
be seen for linear Kihara fluids \cite{vega}.

The studied model fluids have elongations that vary from $L^*=0$, i.e. spherical
molecules, to 0.8, i.e. strongly elongated dumbbell-shaped molecules, in six
steps. The odd value $L^*=0.505$ was chosen to cover the same model fluids 
as the previous work on vapor-liquid equilibria \cite{stoll1}.
Five quadrupole momenta were studied that range from $Q^{*2}=0$ to 4 with
increments of unity. 
The upper limit of 4 is sufficient to describe strongly quadrupolar 
real fluids, eg. CO$_2$ with $Q=-3.7938$ D\r{A} ($Q^{*2}=3.3037$),
C$_2$H$_2$ with $Q=5.0730$ D\r{A} ($Q^{*2}=4$), or
C$_2$F$_4$ with $Q=-7.0332$ D\r{A} ($Q^{*2}=3.9272$) \cite{vrabec}.

For the sake of consistency, transport properties for
spherical fluids ($L^*$=0) were treated as two superimposed Lennard-Jones
sites. This implies, that the temperature has to be divided by 4 as well as the
quadrupole momentum, if a direct comparison with a one-center
Lennard-Jones fluid is to be made. The corresponding conversion of $D^*$,
$\eta^*$, and $\lambda^*$ for these spherical fluids is obtained dividing the
present data by 2.

As temperatures and number densities in vapor-liquid equilibrium 
vary strongly with the molecular parameters, it is useful
to introduce another reduced form for representing the temperature 
$T_R$=$T^*/T_c^*$ and the density $\rho_R$=$\rho^* /\rho_c^*$. 
Here, $T_c^*$ is the critical temperature and $\rho_c^*$ the critical density of the individual
2CLJQ fluid; values for $T_c^*$ and $\rho_c^*$ were taken from \cite{stoll1}. 
For each fluid, the
considered reduced temperatures along the bubble line range from $T_R$=0.6 to 0.9
with increments of $\Delta T_R$=0.1. In addition to those four points, another
12 points in the homogeneous liquid region were simulated, cf. Figs.
\ref{fig1} and \ref{fig2}. These points were selected on isochores starting from
each bubble point with temperature increments of $\Delta T_R$=0.1. In this way,
also isothermal data was generated.

\section{SIMULATION DETAILS}
Equilibrium molecular dynamics simulations were performed in a cubic box of volume
$V$ containing $N=500$ molecules. The cut-off radius was set to $r_{c}=5 \sigma $,
otherwise to half of the box length. The molecules in the fluid were assumed to
have no preferential relative orientations outside of the cut-off sphere. For the
calculation of the LJ long range corrections, orientational averaging was done
with equally weighted relative orientations as proposed by Lustig \cite{lustig}.
The assumption of no preferential relative orientations beyond the cut-off sphere
implies that no long range corrections for the quadrupolar interactions are needed
since they disappear. This is a reasonable assumption as demonstrated by Streett
and Tildesley \cite{strett}. The simulations were started with the molecules in a
face centered cubic lattice with random velocities, the total momentum of the
system was set to zero, and Newton's equations of motion were solved with the Gear
predictor-corrector integration scheme of fifth order \cite{haile}. The time step
for this algorithm was set to $\Delta t \sqrt{\epsilon / m}/\sigma=0.0005$.
Self-diffusion coefficient, shear viscosity, and thermal conductivity were
calculated in the microcanonical $NVE$ ensemble using Eqs. (\ref{self}) to
(\ref{qflux}). The simulations were equilibrated in the canonical $NVT$ ensemble
between $100~000$ to $150~000$ time steps. After equilibration, the thermostat was
turned off and the simulation continued in the $NVE$ ensemble where the transport
properties were calculated. 
Because of the lack of a thermostat, the temperature was fluctuating with a 
maximum drift of 3 \%. 

\begin{figure}[ht]
\caption[Autocorrelation functions (ACF) for self-diffusion coefficient, shear viscosity,
and thermal conductivity for the 2CLJQ fluid with $L^*$=0.2 and $Q^{*2}$=1 at $T^*$=3.0295
and $\rho^*$=0.6573. The vertical line denotes $t^*$=0.1. At this time all normalized
autocorrelation functions have decayed to much less than $1/\textsl{e}$ denoted by the
horizontal lines.]{} \label{fig3}
\begin{center}
\includegraphics[width=150mm,height=200mm]{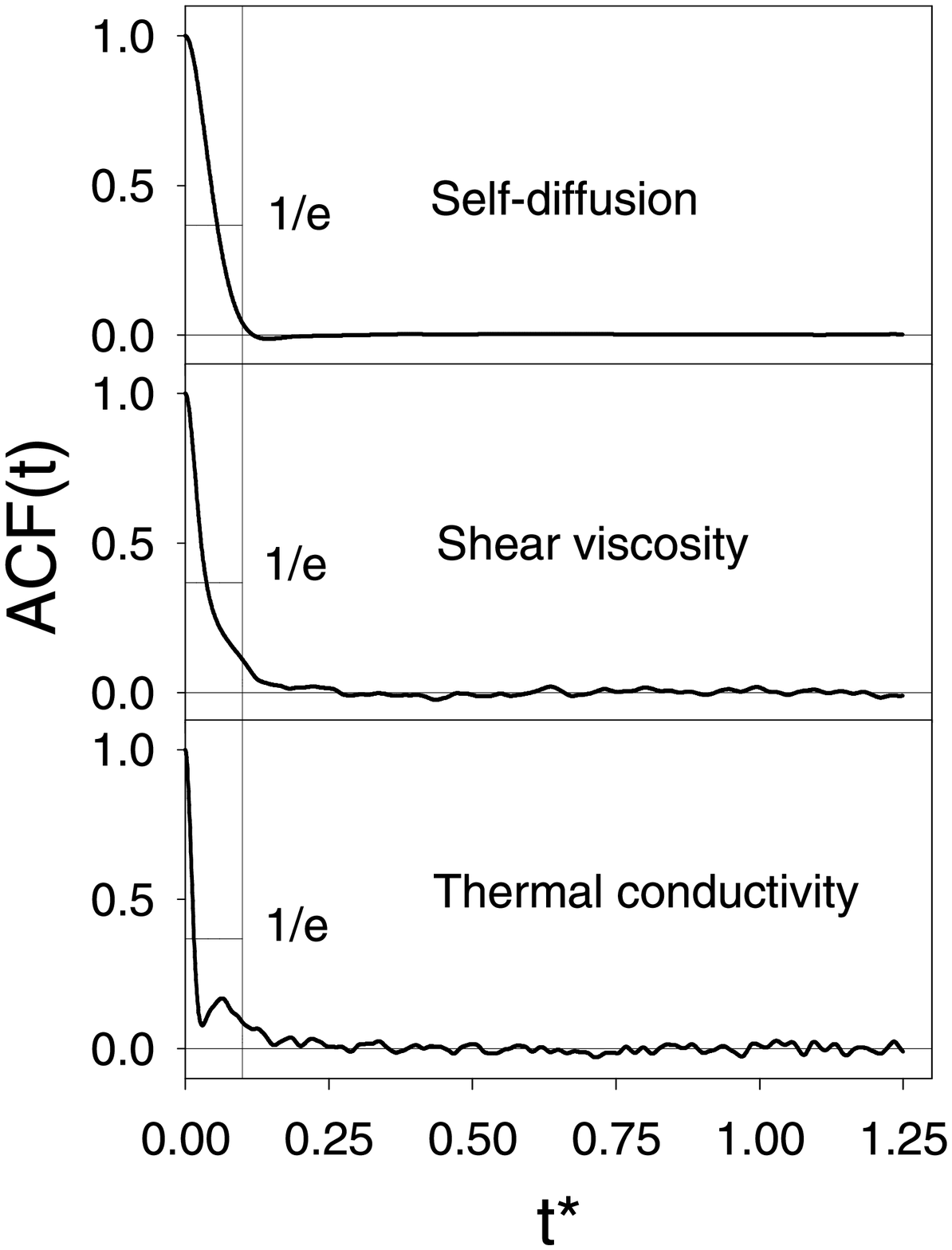}
\end{center}
\end{figure}

\begin{figure}[ht]
\caption[Normalized integrals of autocorrelation functions, i.e. self-diffusion coefficient, shear
viscosity, and thermal conductivity.]{} \label{fig4}
\begin{center}
\includegraphics[width=150mm,height=200mm]{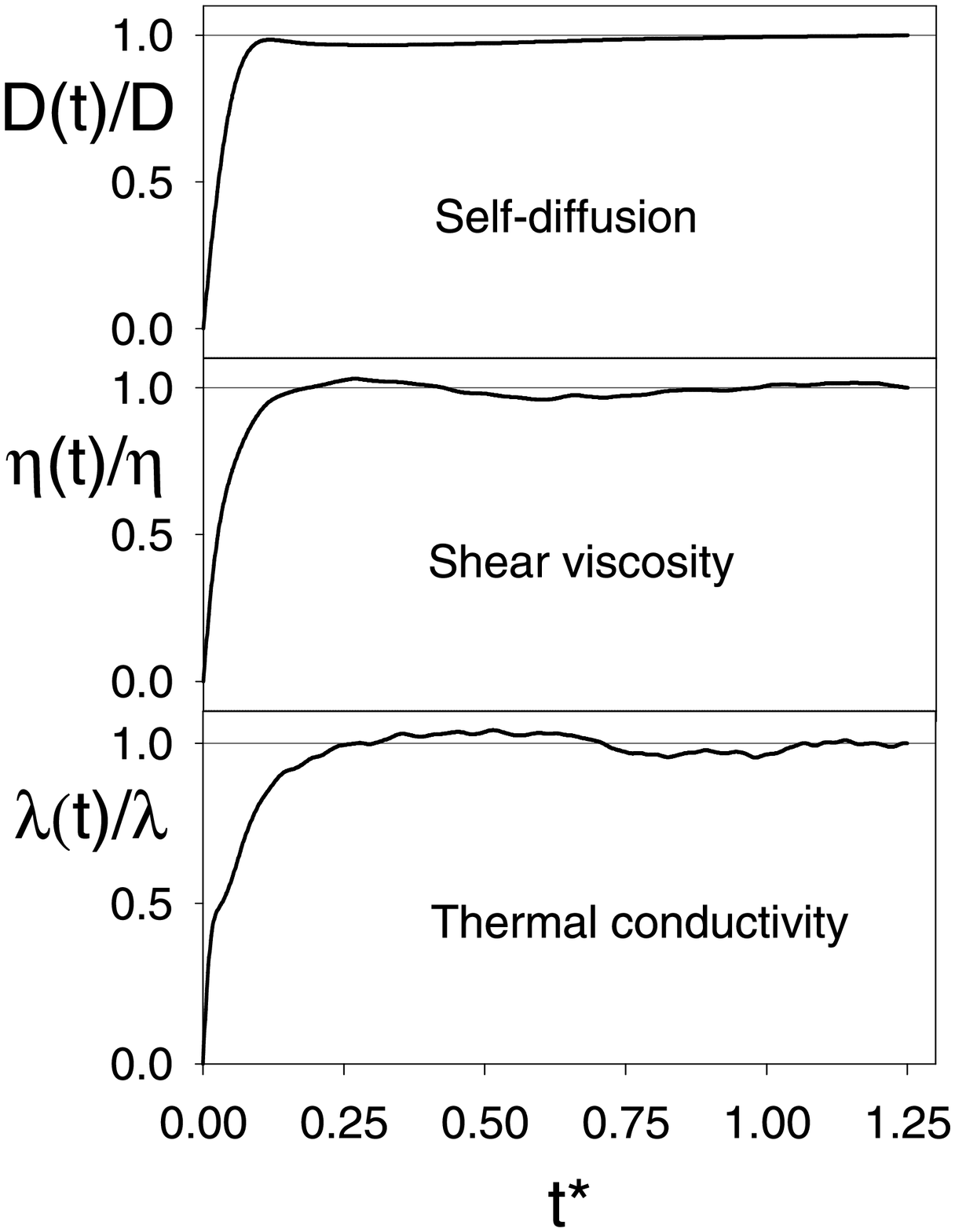}
\end{center}
\end{figure}

Statistical
uncertainties were estimated using the standard deviation of four independent
simulations of $3~000$ independent autocorrelation functions. In order to achieve
independence between autocorrelation functions, a time span of 0.1 in reduced units was left
between consecutive autocorrelation functions. This time span was consistent with
a decay to less than $1/e$ of the normalized velocity autocorrelation function in
several pilot runs. It is a conservative choice, when a compromise between
simulation time and accuracy has to be done. Fig. \ref{fig3} shows the normalized
autocorrelation functions (ACF) from top to bottom for self-diffusion coefficient,
shear viscosity, and thermal conductivity for the 2CLJQ fluid with $L^*$=0.2 and
$Q^{*2}$=1 for the bubble point at 70 \% of its critical temperature
($T_c^*=4.388$), where the bubble density is $\rho^*=0.6573$. The vertical line
denotes $t^*$=0.1, the horizontal line the value $1/e$. All three ACF fulfill the
criterion of independence between consecutive correlations, although they show a
quite different decay. Another important issue is the
significant length of the autocorrelation functions, or equivalently, how long
they must be integrated. Fig. \ref{fig4} shows the integrals from top to bottom
for self-diffusion coefficient, shear viscosity, and thermal conductivity. In
principle, the integration of the ACF must be carried out until the integral shows
stationary behavior. In practice, the convergence of the self-diffusion integral
is difficult to guarantee for all conditions, because of the long time behavior
\cite{alder3} with a decay proportional to $t^{3/2}$. A similar problem is present
regarding the autocorrelation function for shear viscosity close to the
fluid-solid transition \cite{alder2,levesque1,schoen,erpenbeck2}. Here it
was handled by a long evaluation ($\Delta t^*$ = 1.25) of the ACF, as can be seen in Fig.
\ref{fig4}. The integrals converge to the final value at around 0.5,
afterwards the ACF fluctuate around zero, cf. Fig. \ref{fig3}, without effective
contribution to the integrals.

\section{RESULTS}
In this section, the simulation results for the transport coefficients are
presented. Numerical data for self-diffusion coefficient, shear viscosity, and
thermal conductivity are given in Table \ref{tab1a} for elongations from $L^*$=0
to 0.8 and quadrupole momenta from $Q^{*2}$=0 to 4. All data in Table \ref{tab1a}
correspond to state points along the bubble line for reduced temperatures 
of $T_R$=0.6, 0.7, 0.8, and 0.9. The complete data set, with 12
additional state points in the liquid region for each fluid, is available in
\cite{fernandezt} and partially included in Figs. \ref{fig6}, \ref{fig7},
\ref{fig9}, \ref{fig10}, \ref{fig12}, and \ref{fig13}. The effects of elongation,
quadrupole momentum, temperature, and density are discussed in the following for
each transport coefficient separately.

The accuracy of the calculated transport properties decreases in the sequence
self-diffusion coefficient, shear viscosity, thermal conductivity. The high
accuracy of the self-diffusion coefficient, with error bars lower than 3 \%, is due to its
individual nature \cite{hansen}. Shear viscosity and thermal conductivity are
collective properties, consequently they show for the same simulation time and
system size larger uncertainties, that are around 8 and 12 \%, respectively. In most
simulations of the present work the autocorrelation functions of thermal
conductivity decay faster than those for shear viscosity, but fluctuate more.
Figs. \ref{fig3} and \ref{fig4} illustrate this.

Other factors that influence the accuracy of the reported data are elongation and
quadrupole momentum. In particular, at low temperatures, for fluids with large
anisotropy and strong quadrupole momentum, the transport coefficients show larger
simulation uncertainties.

In the following, the results are discussed for nine selected fluids, covering the 
whole range of the two molecular parameters, from spherical ($L^*$=0) over elongated 
($L^*$=0.505) to  strongly elongated ($L^*$=0.8) fluids with varying quadru\-pole 
momentum of $Q^{*2}$=0, 2, and 4. 
A subset of six fluids is taken in some cases only due to graphical reasons. 

\subsection{Self-diffusion coefficient}
\begin{figure}[ht]
\caption[Self-diffusion coefficient of spherical ($L^*$=0, empty symbols), elongated ($L^*$=0.505, grey
symbols), and strongly elongated ($L^*$=0.8, full symbols) 
2CLJQ fluids over reduced density along bubble lines. Reduced temperatures vary from $T_R$=0.6 to
0.9. Lines are guides for the eye.]{} \label{fig5a}
\begin{center}
\includegraphics[width=150mm,height=200mm]{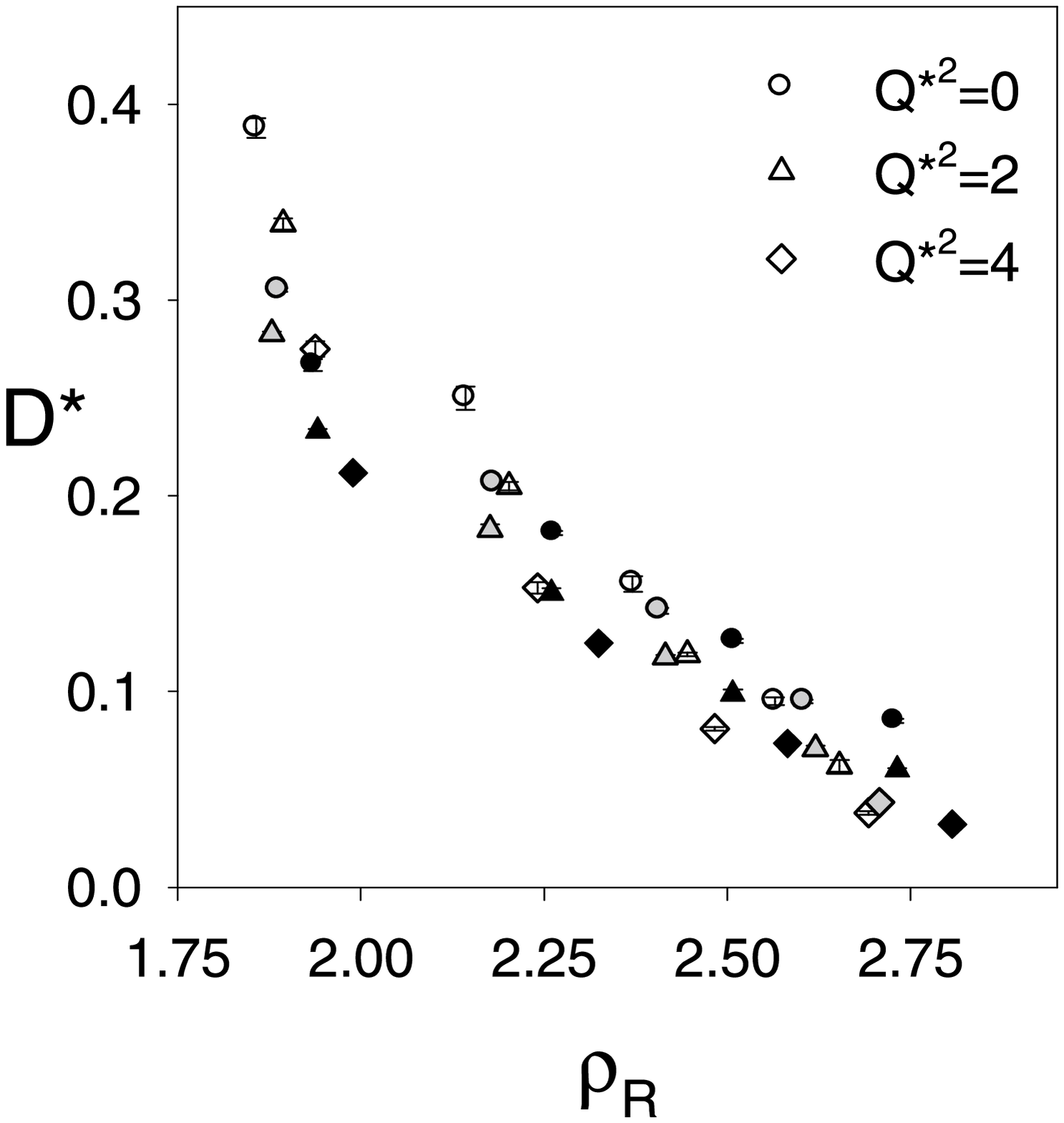}
\end{center}
\end{figure}

\begin{figure}[ht]
\caption[Self-diffusion coefficient of spherical ($L^*$=0, empty symbols), elongated ($L^*$=0.505, grey
symbols), and strongly elongated ($L^*$=0.8, full symbols) 2CLJQ fluids over number density along bubble lines. Reduced temperatures vary from $T_R$=0.6 to
0.9. Lines are guides for the eye.]{} \label{fig5}
\begin{center}
\includegraphics[width=150mm,height=200mm]{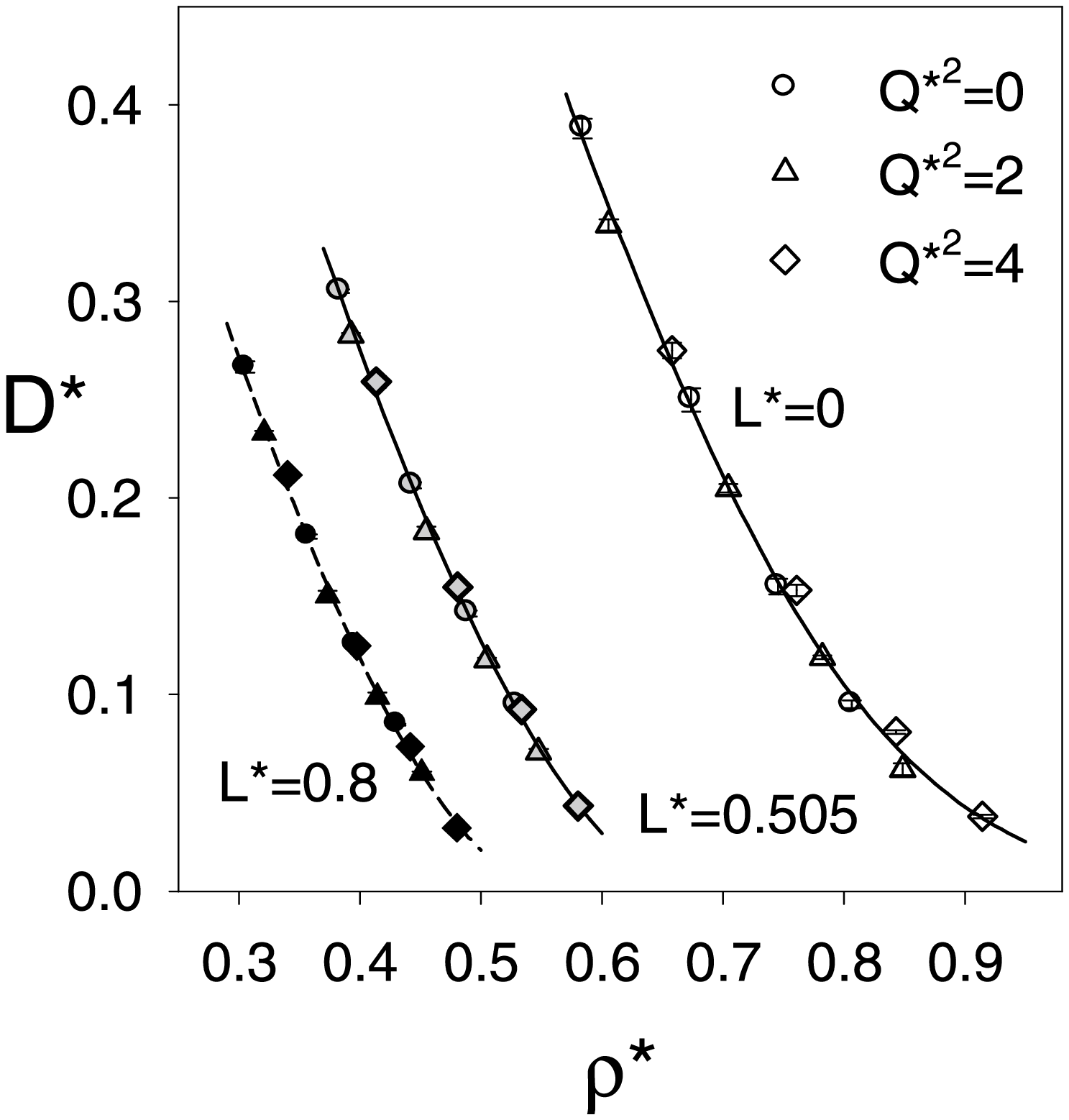}
\end{center}
\end{figure}

\begin{figure}[ht]
\caption[Self-diffusion coefficient of spherical ($L^*$=0, empty symbols), elongated ($L^*$=0.505, grey
symbols), and strongly elongated ($L^*$=0.8, full symbols) 2CLJQ fluids over 
number density in the homogeneous liquid at $T_R$=0.9. Lines are guides for the eye.]{} \label{fig6}
\begin{center}
\includegraphics[width=150mm,height=200mm]{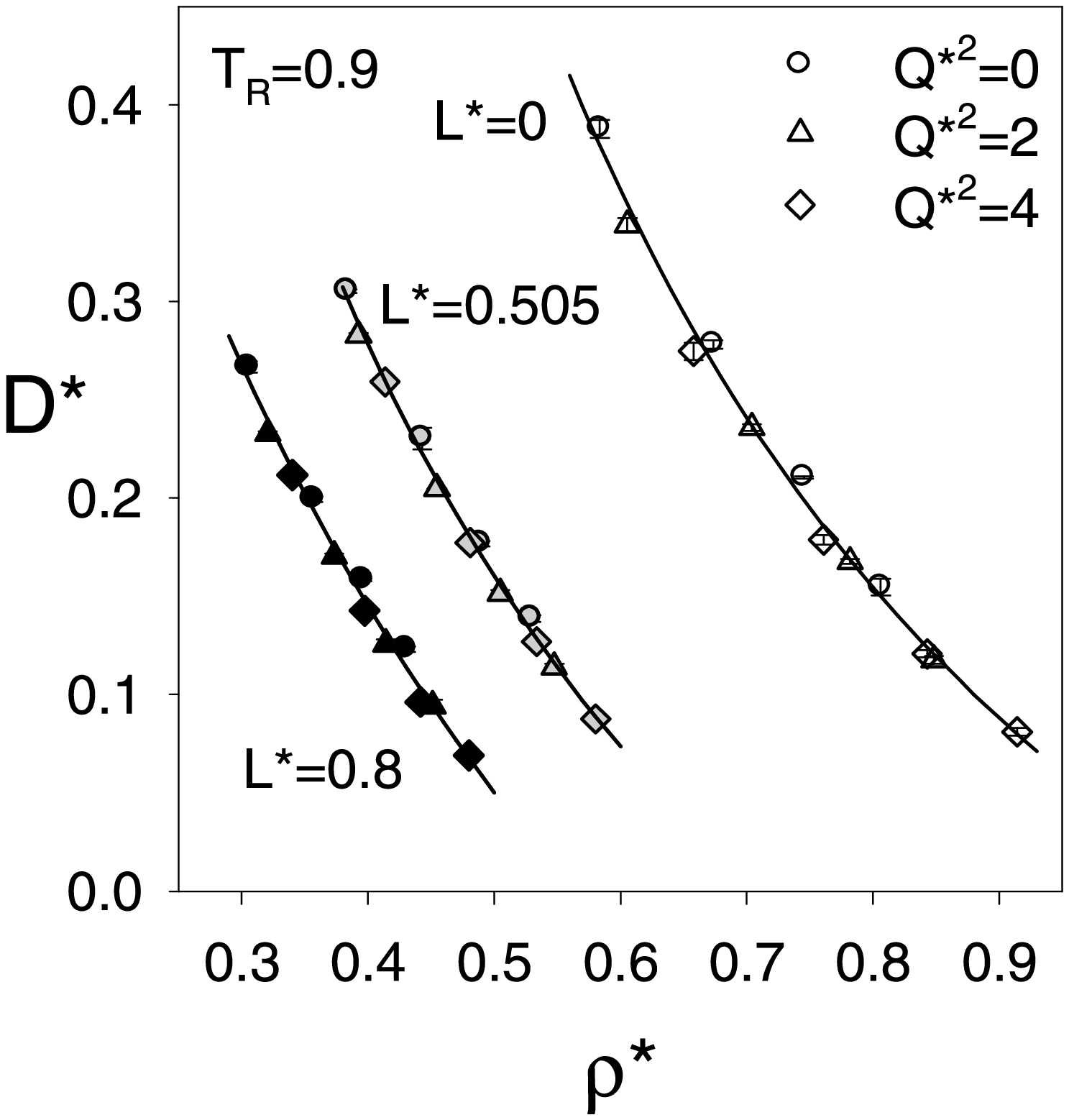}
\end{center}
\end{figure}

\begin{figure}[ht]
\caption[Self-diffusion coefficient of spherical ($L^*$=0, empty symbols) and strongly elongated ($L^*$=0.8, full
symbols) 2CLJQ fluids over reduced temperature 
in the homogeneous liquid along different isochores. $\circ$: $\rho^*$=0.8062,
$\vartriangle$: $\rho^*$=0.8483, $\lozenge$: $\rho^*$=0.9143,
$\bullet$: $\rho^*$=0.4302, $\blacktriangle$: $\rho^*$=0.4513, 
$\blacklozenge$: $\rho^*$=0.4800. 
Lines are guides for the eye.]{}
\label{fig7}
\begin{center}
\includegraphics[width=150mm,height=200mm]{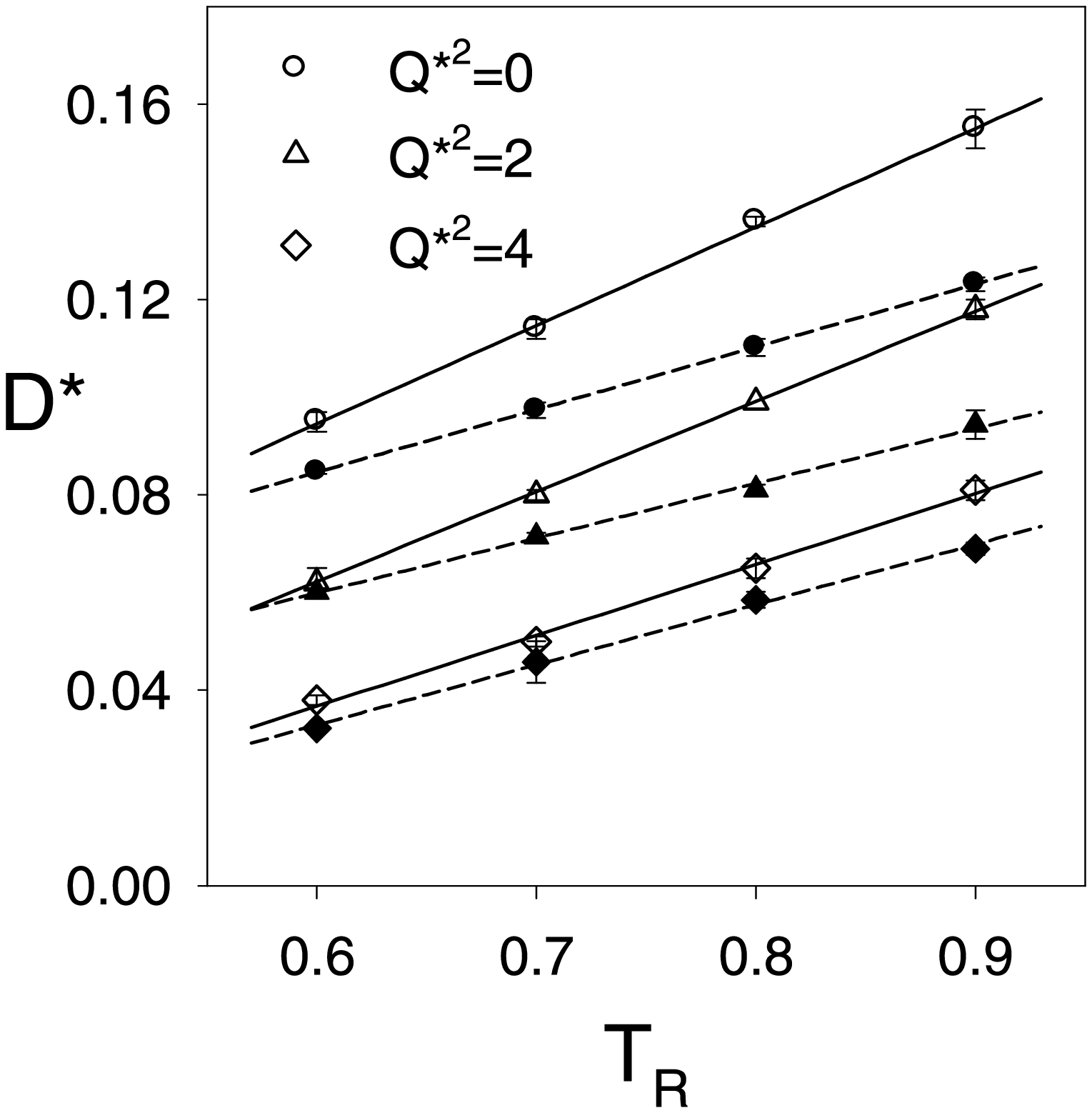}
\end{center}
\end{figure}

Figs. \ref{fig5a} and \ref{fig5} illustrate the self-diffusion coefficient along 
the bubble line for nine selected fluids. 
The results can either be discussed in terms of reduced density $\rho_R$ 
as in Fig. \ref{fig5a} or in terms of number density $\rho^*$ as in Fig. \ref{fig5}.
From Fig. \ref{fig5a} it can be seen that the regarded range of reduced density is similar 
for all fluids, but significant deviations from the principle of corresponding states 
are present also for the density. 
At constant $T_R$, it can be discerned that the self-diffusion
coefficient decreases with both increasing elongation and quadrupole momentum. 
A better visibility of the data (which is even more needed for the less accurate properties 
shear viscosity and thermal conductivity) is obtained when plotted over number density in 
Fig. \ref{fig5}. Therefore, this graphical representation is preferred in the following.

As Fig. \ref{fig5} shows, $D^*$ decreases with increasing number density along the bubble line
(where with increasing density also the temperature decreases). It is an important
result of the present study that the self-diffusion coefficient lies 
roughly along the same line for a given elongation, independent of the quadrupole momentum. 

Fig. \ref{fig6} shows the dependence of $D^*$ on number density 
in the homogeneous liquid region at a constant reduced
temperature of $T_R$=0.9. Note that the density range is the same as in Fig.
\ref{fig5}. Along this isotherm $D^*$ decreases slightly hyperbolic 
with increasing density,
resembling the behavior of $D^*$ along bubble lines for a given elongation.
Comparing $D^*$ along bubble lines with isothermal data for an identical density
variation, it is found that the density effect dominates with a contribution of 80 \%. 

Fig. \ref{fig7} shows the dependence of the self-diffusion coefficient on
temperature at different constant densities for a subset of six fluids. 
The isochores correspond to bubble densities at the reduced
temperature $T_R$=0.6, cf. Figs. \ref{fig1} and \ref{fig2}, which have similar values 
in terms of $\rho_R$. Along the isochores, the
self-diffusion coefficient increases linearly with increasing temperature. The
gradients with respect to reduced temperature are almost constant for a given
elongation, where the slope is less steep for more elongated fluids. Such a linear
dependence of $D^*$ on temperature has also been reported by other authors for
Lennard-Jones fluids \cite{levesque2}, Kihara fluids \cite{macdowell}, and
two-center Lennard-Jones fluids \cite{singer}.

\subsection{Shear viscosity}
\begin{figure}[ht]
\caption[Shear viscosity of spherical ($L^*$=0, empty symbols), elongated ($L^*$=0.505, grey
symbols), and strongly elongated ($L^*$=0.8, full symbols)  2CLJQ fluids over 
number density along bubble lines. Reduced temperatures vary from $T_R$=0.6 to 0.9.
Lines are guides for the eye.]{} \label{fig8}
\begin{center}
\includegraphics[width=150mm,height=200mm]{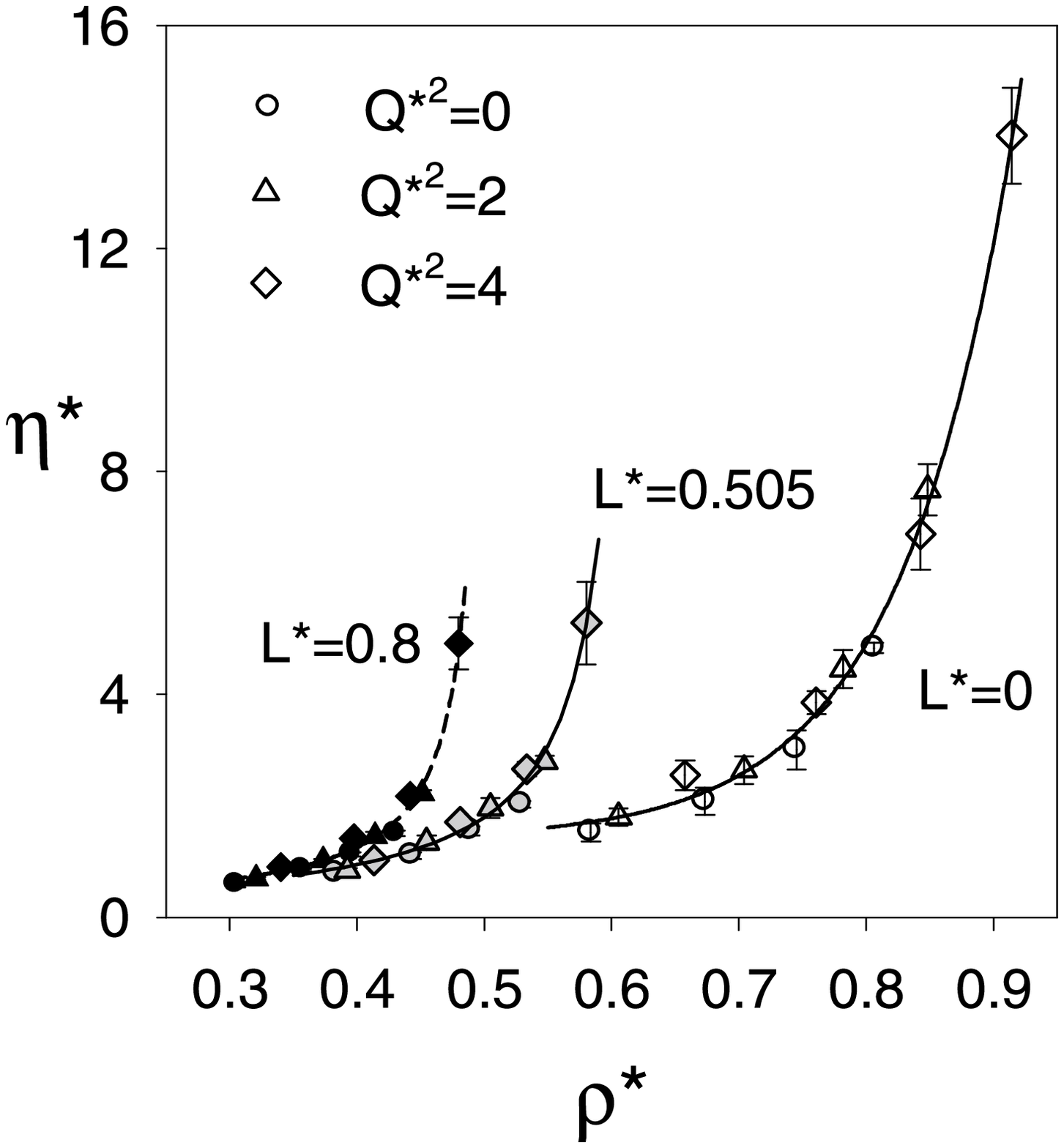}
\end{center}
\end{figure}

\begin{figure}[ht]
\caption[Shear viscosity of spherical ($L^*$=0, empty symbols), elongated ($L^*$=0.505, grey
symbols), and strongly elongated ($L^*$=0.8, full symbols) 2CLJQ fluids over 
number density in the homogeneous liquid
at $T_R$=0.9. Lines are guides for the eye.]{} \label{fig9}
\begin{center}
\includegraphics[width=150mm,height=200mm]{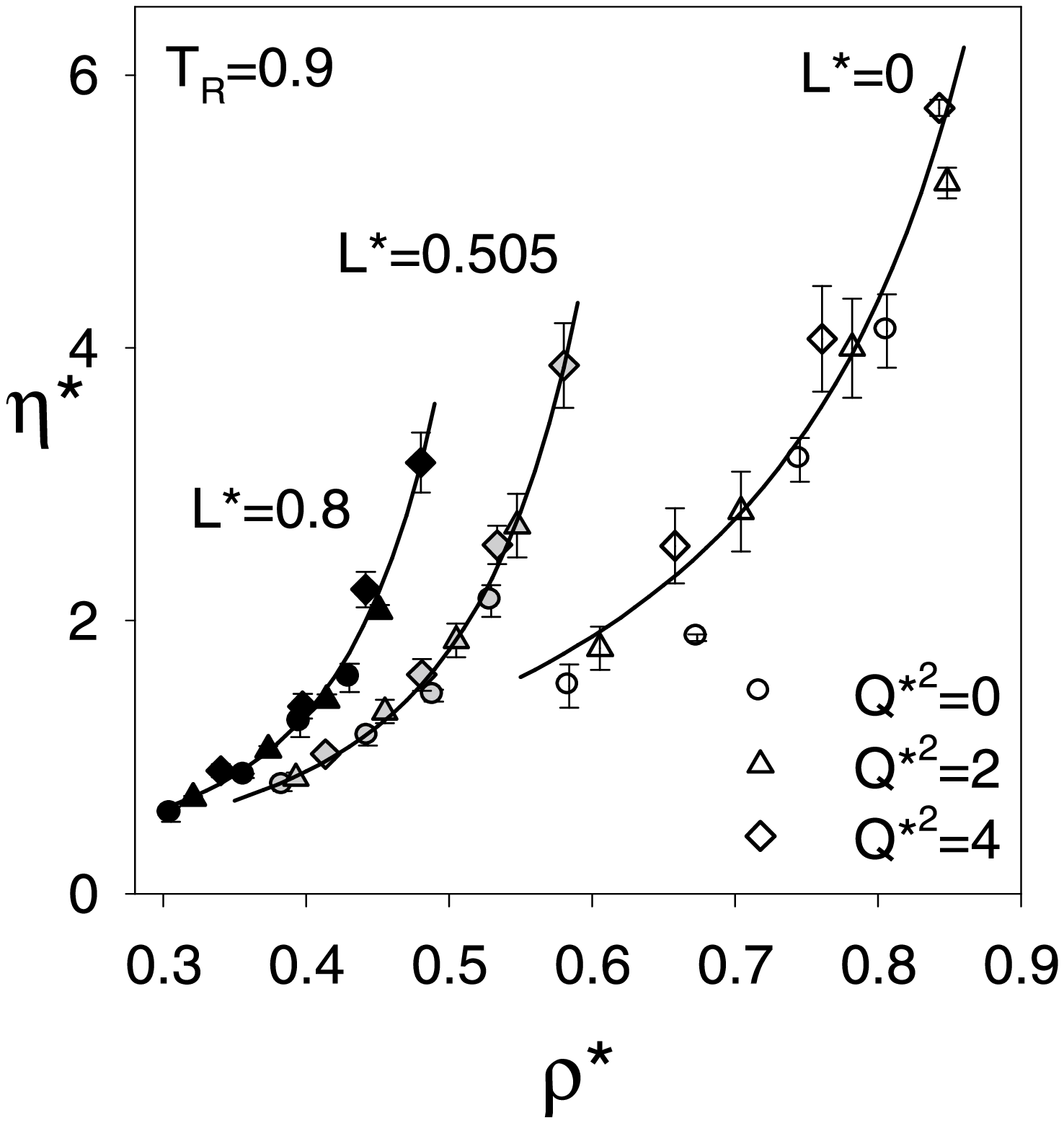}
\end{center}
\end{figure}

\begin{figure}[ht]
\caption[Shear viscosity of spherical ($L^*$=0, empty symbols) and strongly elongated ($L^*$=0.8, full
symbols) 2CLJQ fluids over reduced temperature 
in the homogeneous liquid along different isochores. $\circ$: $\rho^*$=0.8062,
$\vartriangle$: $\rho^*$=0.8483, $\lozenge$: $\rho^*$=0.9143, 
$\bullet$: $\rho^*$=0.4302, $\blacktriangle$: $\rho^*$=0.4513, 
$\blacklozenge$: $\rho^*$=0.4800. Lines are guides for the eye.]{}
\label{fig10}
\begin{center}
\includegraphics[width=150mm,height=200mm]{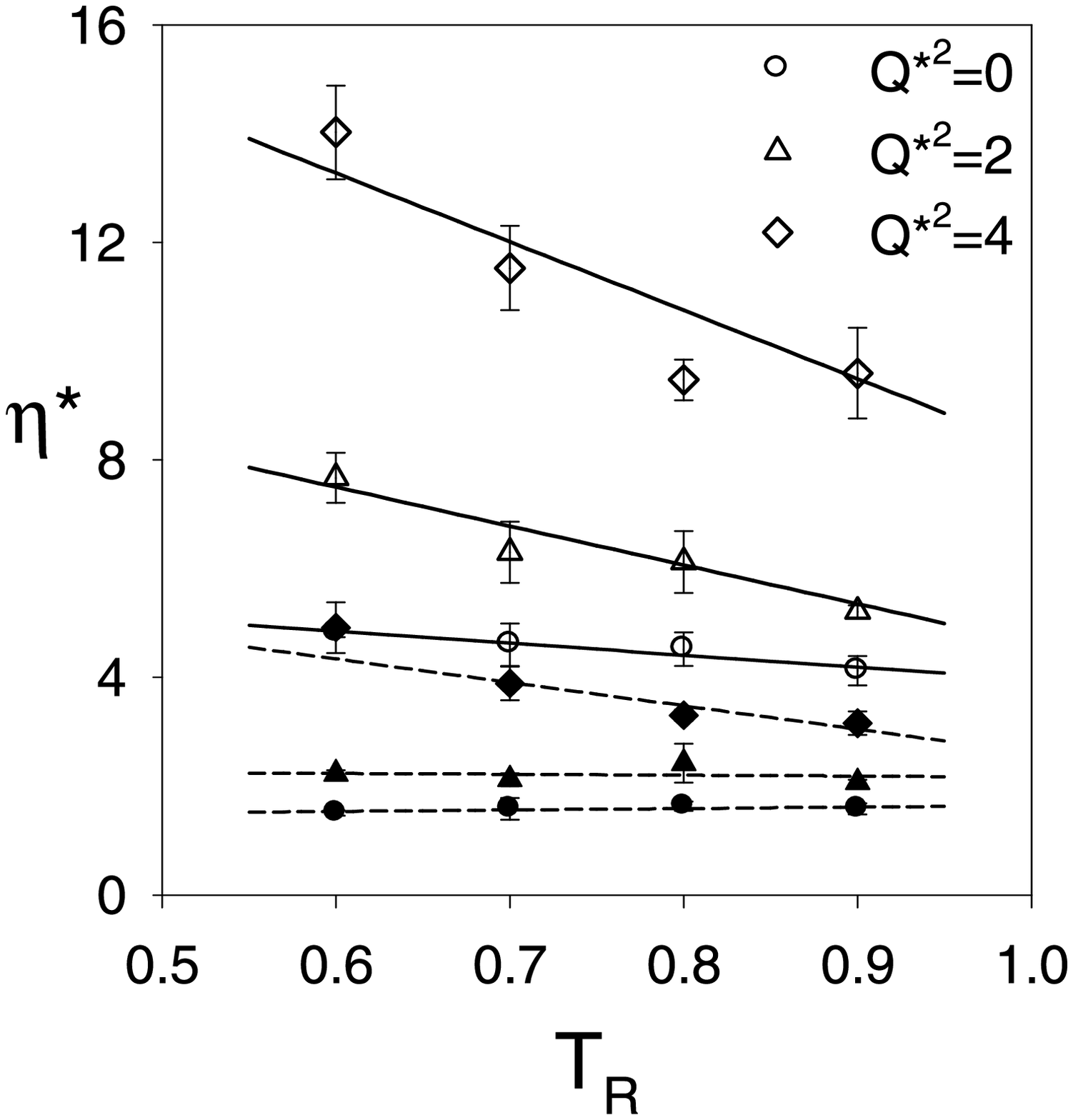}
\end{center}
\end{figure}

Fig. \ref{fig8} illustrates the shear viscosity along the bubble line for the nine selected fluids.
At constant $T_R$, it is found that the shear viscosity decreases with increasing elongation 
but increases with increasing quadrupole momentum. 
Again it is found that the results for a given elongation lie roughly along one line
independent of $Q^{*2}$, where, as expected, they increase with increasing number density. 

The density dependence of shear viscosity in the homogeneous liquid region 
is illustrated in Fig. \ref{fig9} at $T_R$=0.9. 
Comparing the variation of $\eta^*$ along bubble lines and along
isotherms in the same way as for $D^*$, it is found for non-polar fluids that the
density effect is responsible for about 80 \% of the increase of $\eta^*$ along
the bubble line. For quadrupolar fluids, however, the temperature influence
becomes more important and its contribution is about 40 \%.

Fig. \ref{fig10} shows the dependence of shear viscosity on reduced temperature
for a subset of six fluids along isochores with similar values in terms of $\rho_R$. 
As expected, shear viscosity decreases with increasing temperature. 
Strongly quadrupolar fluids, with an about threefold higher shear viscosity
in the cold liquid, are more sensitive to temperature, exhibiting larger gradients.

\subsection{Thermal conductivity}
\begin{figure}[ht]
\caption[Thermal conductivity of spherical ($L^*$=0, empty symbols), elongated ($L^*$=0.505, grey
symbols), and strongly elongated ($L^*$=0.8, full symbols) 2CLJQ fluids over 
number density along bubble lines. 
Reduced temperatures vary from $T_R$=0.6 to 0.9.
Lines are guides for the eye.]{} \label{fig11}
\begin{center}
\includegraphics[width=150mm,height=200mm]{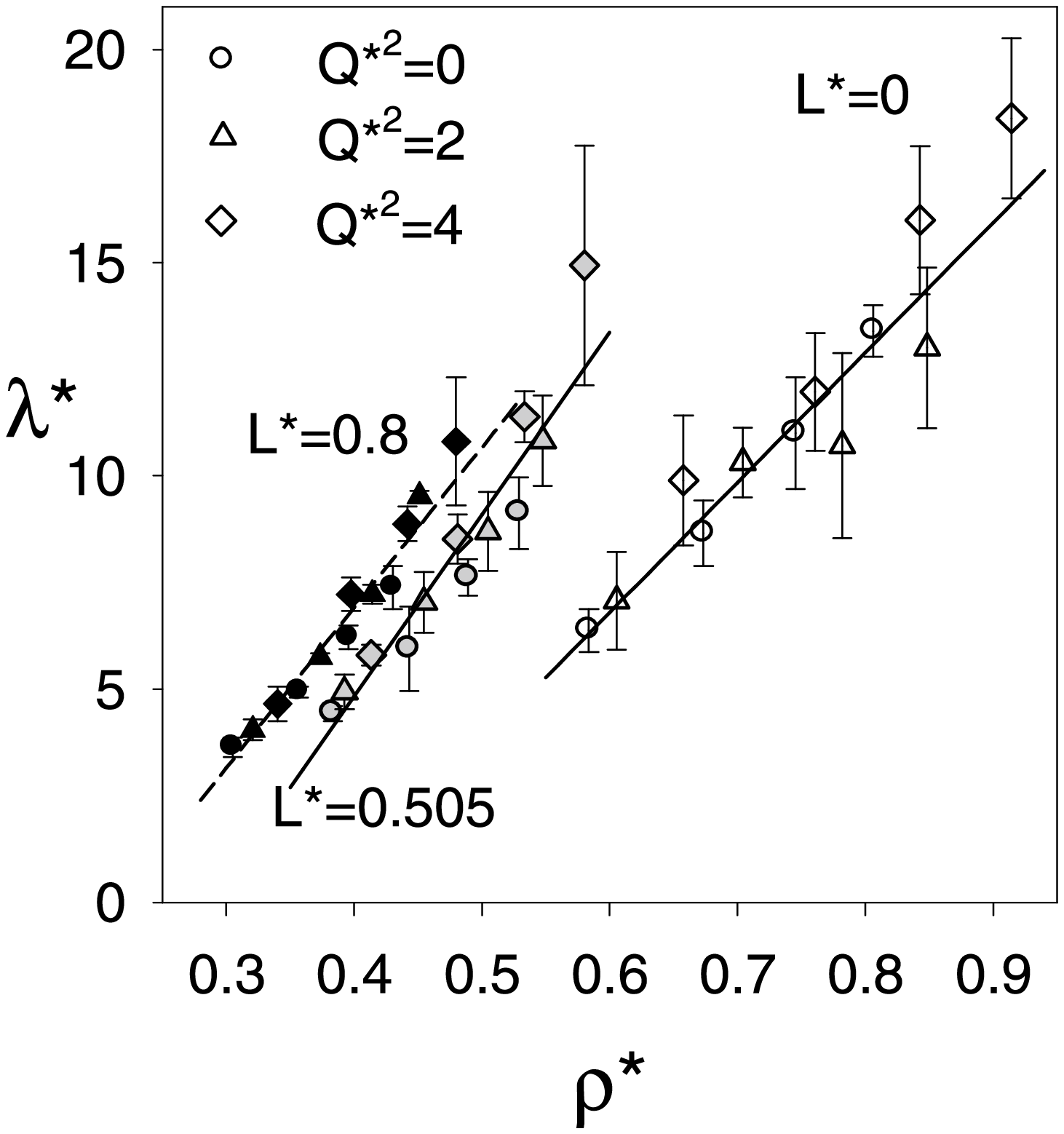}
\end{center}
\end{figure}

\begin{figure}[ht]
\caption[Thermal conductivity of spherical ($L^*$=0, empty symbols), elongated ($L^*$=0.505, grey
symbols), and strongly elongated ($L^*$=0.8, full symbols) 2CLJQ fluids over 
number density in the homogeneous liquid at $T_R$=0.9. 
Lines are guides for the eye.]{} \label{fig12}
\begin{center}
\includegraphics[width=150mm,height=200mm]{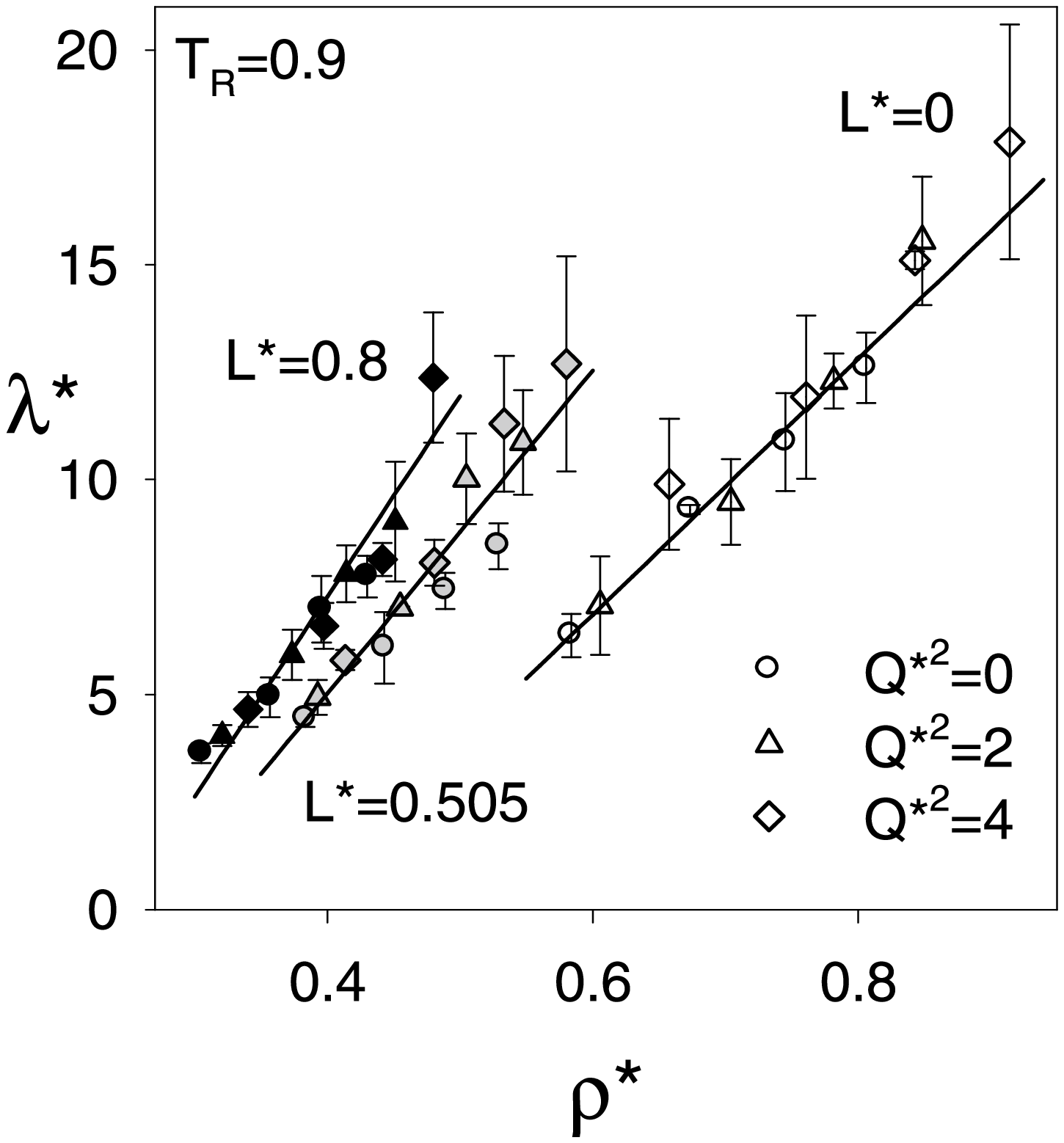}
\end{center}
\end{figure}

\begin{figure}[ht]
\caption[Thermal conductivity of spherical ($L^*$=0, empty symbols) and strongly elongated ($L^*$=0.8, full
symbols) 2CLJQ fluids over reduced temperature 
in the homogeneous liquid along different isochores. $\circ$: $\rho^*$=0.8062,
$\vartriangle$: $\rho^*$=0.8483, $\lozenge$: $\rho^*$=0.9143,
$\bullet$: $\rho^*$=0.4302, $\blacktriangle$: $\rho^*$=0.4513, 
$\blacklozenge$: $\rho^*$=0.4800. 
Lines are guides for the eye.]{}
\label{fig13}
\begin{center}
\includegraphics[width=150mm,height=200mm]{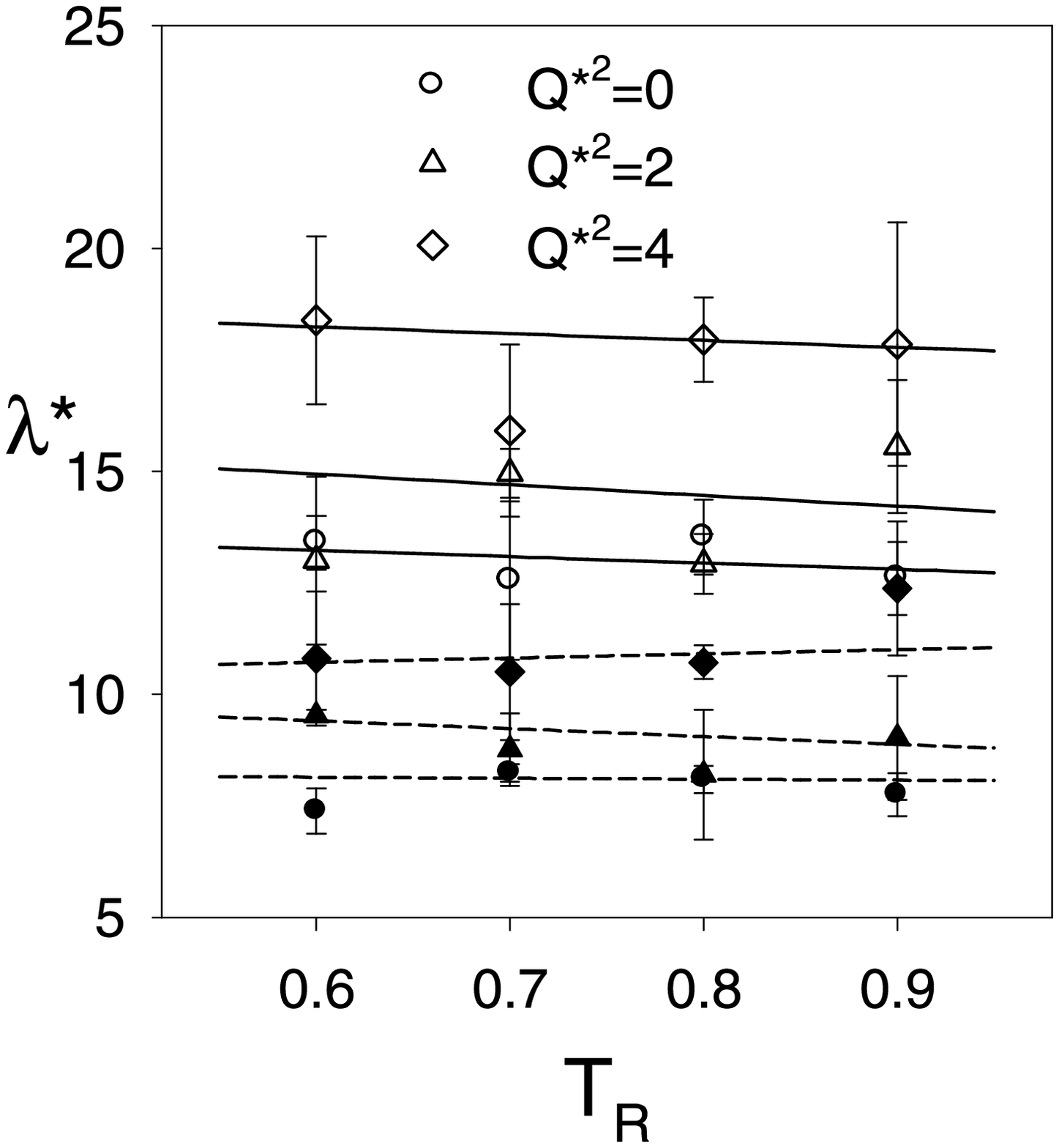}
\end{center}
\end{figure}

Fig. \ref{fig11} illustrates the thermal conductivity along the bubble line.
Again, the data lie roughly along single
lines for a given elongation, but considering simulation uncertainties not more than a
linear dependence can be discerned. 
Thermal conductivity has the same basic trends like shear viscosity
as it decreases with increasing elongation but increases with increasing quadrupole momentum
at constant $T_R$. 

Fig. \ref{fig12} shows the density dependence in the homogeneous liquid
at $T_R$=0.9. It can be seen that the curves resemble those 
along bubble lines, underlining the dominating effect of density there. Similar results
have been reported by Tokumasu et al. \cite{tokumasu} who studied the non-polar
2CLJ potential but at different thermodynamic conditions. In their analysis,
Tokumasu et al. reduced $\lambda^*$ by critical temperature and critical density, to
isolate the effect of elongation and found that this type of reduced thermal
conductivity increases with increasing elongation.

Fig. \ref{fig13} shows isochoric data with similar values in terms of $\rho_R$ for a 
subset of six fluids, where the effect of temperature on $\lambda^*$ is small.
Taking the statistical uncertainty and the scatter into account, hardly any trend
can be discerned. Experimental results \cite{slyusar1,slyusar2} show that thermal
conductivity at constant density increases with increasing temperature, but the
variation is very small in the liquid region. Moreover, the increase of $\lambda^*$
with increasing quadrupole momentum can be seen.

\section{CONCLUSION}
Equilibrium molecular dynamics simulation and the Green-Kubo
formalism were used to calculate self-diffusion coefficient, shear viscosity, and
thermal conductivity for 30 different anisotropic and quadrupolar model fluids. 
A comprehensive data set was obtained for each fluid and property that covers 
a substantial part of the liquid state. 
The statistical uncertainty of the reported data varies according to
transport property. For self-diffusion coefficient data, it is less than 3 \%,
for shear viscosity and thermal conductivity it is around 8 and 12 \%,
respectively.

The three transport properties are dominated in the investigated liquid region 
by the density: saturated liquid and isothermal
data for fluids with a given elongation but varying quadrupole momentum 
lie roughly along single lines when plotted over number density.

However, all transport properties on the bubble line at a constant reduced temperature 
are lower for fluids with larger elongation. 
An increasing quadrupole momentum also leads to a lower
self-diffusion coefficient, the opposite is found for shear viscosity and 
thermal conductivity. 

Temperature influences all transport properties less than density. As expected,
along isochores, the self-diffusion coefficient increases with temperature, the shear
viscosity decreases, and for the thermal conductivity hardly any variation can
be discerned.

\clearpage

\noindent
{\bf List of symbols}

\begin{tabular}{ll}
$a$ & interaction site index \\
$b$ & interaction site index \\
$c$ & short notation for a trigonometric function \\
$D$ & self-diffusion coefficient \\
$i$ & molecule index \\
$j$ & molecule index \\
$J_p$ & element of the microscopic stress tensor \\
$J_q$ & element of the microscopic heat flow \\
$k$ & interaction site index \\
$k$ & molecule index \\
$k_B$ & Boltzmann constant \\
$l$ & interaction site index \\
$L$ & molecular elongation \\
$m$ & molecular mass \\
$N$ & number of molecules \\
$Q$ & molecular quadrupole momentum \\
$r$ & site-site distance \\
$r_{\rm c}$ & center-center cut-off radius \\
$s$ & short notation for a trigonometric function \\
$t$ & time \\
$T$ & temperature \\
$u$ & pair potential \\
$v$ & element of the velocity vector \\
$V$ & molar volume \\
$\Delta$ & increment \\
$\Delta t$ & integration time step\\
$\epsilon$ & Lennard-Jones energy parameter \\
$\epsilon_0$ & Electric constant \\
\end{tabular}

\begin{tabular}{ll}
$\eta$ & shear viscosity \\
$\theta$  & angle of nutation \\
$\lambda$ & thermal conductivity \\
$\rho$    & density \\
$\sigma$  & Lennard-Jones size parameter \\
$\phi$    & azimuthal angle \\
\end{tabular}

\noindent
\textbf{Vector properties} \\[0.1cm]
\begin{tabular}{ll}
$\textbf{I}$ & matrix of angular momentum of inertia \\
$\textbf{J}_p$ & microscopic stress tensor \\
$\textbf{J}_q$ & microscopic heat flow vector \\
$\textbf{r}$ & distance vector \\
$\textbf{v}$ & velocity vector \\
$\textbf{w}$ & angular velocity vector \\
$\mathbf{\Gamma}$ & torque vector \\
\end{tabular}

\noindent
\textbf{Subscript} \\[0.1cm]
\begin{tabular}{ll}
$a$ & interaction site index \\
$b$ & interaction site index \\
$c$ & property at critical point \\
$i$ & molecule index \\
$j$ & molecule index \\
Q & point quadrupole \\
$R$ & property reduced by critical value \\
2CLJQ & two-center Lennard-Jones plus point quadrupole \\
\end{tabular}

\noindent
\textbf{Superscript} \\[0.1cm]
\begin{tabular}{ll}
$x$ & cartesian direction \\
$y$ & cartesian direction \\
$z$ & cartesian direction \\
*   & property reduced by molecular parameters \\
\end{tabular}

\clearpage

\noindent 
{\bf Acknowledgement}

We gratefully acknowledge financial support from Deutscher Akademischer Austauschdienst (DAAD).

\clearpage

\clearpage
\begin{table}[t]
\caption{Transport coefficients along the bubble line for 30 2CLJQ fluids of different elongation $L^*$ and
quadrupole momentum $Q^{*2}$. The numbers in parentheses denote the uncertainty in the last digits.}\label{tab1a}
\bigskip
\begin{center}
\begin{tabular}
{lcc|ccc}
\hline\hline
$L^*$=0 & $T^*$ & $\rho^*$ & $D^*$ & $\eta^*$ & $\lambda^*$  \\
\hline\hline

$Q^{*2}$=0 & 3.156  & 0.8062 & 0.095(2) & \04.83(9)\0 & 13.4(6)\0\0   \\
           & 3.681  & 0.7453 & 0.155(4) & \03.00(35)  & 11.0(13)\0    \\
           & 4.255  & 0.6735 & 0.250(6) & \02.00(24)  & \08.65(77)  \\
           & 4.674  & 0.5838 & 0.388(5) & \01.52(16)  & \06.37(51)  \\
\hline
$Q^{*2}$=1 & 3.132  & 0.8201 & 0.080(1) & \05.31(35) & 11.7(12)\0 \\
           & 3.712  & 0.7575 & 0.143(2) & \03.62(23) & 11.2(12)\0 \\
           & 4.187  & 0.6851 & 0.228(4) & \02.26(18) & \09.8(12)\0  \\
           & 4.773  & 0.5934 & 0.368(2) & \01.58(6)\0  & \06.44(78)  \\
\hline
$Q^{*2}$=2 & 3.398  & 0.8483 & 0.062(3) & \07.67(46) & 13.0(19)\0 \\
           & 3.925  & 0.7819 & 0.119(1) & \04.45(34) & 10.7(22)\0 \\
           & 4.538  & 0.7041 & 0.205(2) & \02.64(25) & 10.32(82) \\
           & 4.946  & 0.6055 & 0.339(3) & \01.79(16) & \07.07(11)\\
\hline
$Q^{*2}$=3 & 3.505  & 0.8796 & 0.047(1) & \09.70(64) & 15.4(25)\0 \\
           & 4.106  & 0.8099 & 0.099(2) & \05.83(58) & 12.4(13)\0 \\
           & 4.741  & 0.7292 & 0.179(1) & \03.28(62) & 10.4(19)\0 \\
           & 5.341  & 0.6272 & 0.310(4) & \01.93(26) & \07.9(12)\0  \\
\hline
$Q^{*2}$=4 & 3.841  & 0.9143   & 0.038(1)  & 14.02(86)  & 18.3(19)\0 \\
           & 4.476  & 0.8430   & 0.081(1)  & \06.87(64)   & 15.9(17)\0 \\
           & 5.164  & 0.7609   & 0.153(3)  & \03.85(21)   & 11.9(14)\0 \\
           & 5.742  & 0.6579   & 0.275(4)  & \02.54(27)   & \09.8(15)\0  \\
\hline\hline
\end{tabular}
\end{center}
\end{table}

\setcounter{table}{0}
\begin{table}[t]
\noindent \caption{Continued.}
\bigskip
\begin{center}
\begin{tabular}
{lcc|ccc} \hline\hline
$L^*$=0.2 & $T^*$ & $\rho^*$ & $D^*$ & $\eta^*$ & $\lambda^*$  \\
\hline\hline
$Q^{*2}$=0 &  2.589 &  0.7114 & 0.089(1) & \03.64(16)   &11.58(59)\\
           &  3.015 &  0.6573 & 0.147(1) & \02.63(7)\0  &\09.69(74) \\
           &  3.441 &  0.5946 & 0.228(6) & \01.74(11)   &\07.67(83) \\
           &  3.874 &  0.5144 & 0.353(7) & \01.27(5)\0  &\05.90(30) \\
\hline
$Q^{*2}$=1 &  2.625 &  0.7203 & 0.083(1) & \04.19(21)   &12.1(10)\0 \\
           &  3.070 &  0.6644 & 0.142(2) & \02.86(23)   &11.7(13)\0 \\
           &  3.496 &  0.5998 & 0.226(4) & \01.89(9)\0  &\08.54(57)\\
           &  3.941 &  0.5185 & 0.350(9) & \01.19(7)\0  &\06.74(28)\\
\hline
$Q^{*2}$=2 &  2.722 &  0.7420 & 0.075(1) & \04.78(36)   &17.1(14)\0 \\
           &  3.195 &  0.6833 & 0.133(2) & \02.97(12)   &15.4(15)\0 \\
           &  3.659 &  0.6167 & 0.215(4) & \02.12(14)   &11.76(93) \\
           &  4.072 &  0.5322 & 0.338(8) & \01.34(8)\0  &\07.74(47) \\
\hline
$Q^{*2}$=3 &  2.877 &  0.7683 & 0.067(1) & \05.48(21)    &21.0(29)\0  \\
           &  3.393 &  0.7085 & 0.123(1) & \03.67(3)\0   &19.8(16)\0  \\
           &  3.856 &  0.6397 & 0.203(5) & \02.40(7)\0   &15.6(18)\0  \\
           &  4.318 &  0.5535 & 0.326(6) & \01.53(16)    &\09.51(10) \\
\hline
$Q^{*2}$=4 &  3.054 &  0.7939 & 0.026(3) & 11.21(42)     &23.3(31)\0 \\
           &  3.642 &  0.7311 & 0.117(2) & \04.06(14)    &22.5(24)\0 \\
           &  4.103 &  0.6596 & 0.196(2) & \02.44(8)\0   &18.4(21)\0 \\
           &  4.600 &  0.5686 & 0.324(10)& \01.71(10)    &12.0(21)\0 \\
\hline\hline
\end{tabular}
\end{center}
\end{table}

\setcounter{table}{0}
\begin{table}[t]
\noindent \caption{Continued.}
\bigskip
\begin{center}
\begin{tabular}
{lcc|ccc} \hline\hline
$L^*$=0.4 & $T^*$ & $\rho^*$ & $D^*$ & $\eta^*$ & $\lambda^*$  \\
\hline\hline
$Q^{*2}$=0 & 1.893  & 0.5808 & 0.094(2) &  2.51(20)   &\09.36(45) \\
           & 2.232  & 0.5365 & 0.147(1) &  1.82(9)\0  &\08.90(36) \\
           & 2.536  & 0.4853 & 0.219(5) &  1.32(9)\0  &\06.67(57) \\
           & 2.858  & 0.4185 & 0.326(3) &  0.89(5)\0  &\04.92(47) \\
\hline
$Q^{*2}$=1 & 1.925  & 0.5879 & 0.084(1) &  2.82(17)   &10.12(60)  \\
           & 2.239  & 0.5426 & 0.134(1) &  2.03(7)\0  &\08.97(38) \\
           & 2.573  & 0.4913 & 0.206(1) &  1.35(5)\0  &\07.39(55) \\
           & 2.869  & 0.4252 & 0.311(2) &  0.92(8)\0  &\05.24(84) \\
\hline
$Q^{*2}$=2 & 1.999  & 0.6025 & 0.072(1) &  3.36(11)   &11.11(78)  \\
           & 2.318  & 0.5555 & 0.121(1) &  2.23(6)\0  &\09.58(33) \\
           & 2.663  & 0.5008 & 0.194(2) &  1.53(8)\0  &\07.60(55) \\
           & 2.987  & 0.431  & 0.304(4) &  0.97(5)\0  &\06.02(37) \\
\hline
$Q^{*2}$=3 & 2.096  & 0.6209 & 0.061(1) &  3.98(14)   &14.9(11)\0 \\
           & 2.447  & 0.5717 & 0.109(1) &  2.64(3)\0  &10.52(82)  \\
           & 2.794  & 0.5154 & 0.180(1) &  1.71(7)\0  &\08.52(38) \\
           & 3.134  & 0.4428 & 0.290(5) &  1.07(6)\0  &\06.37(56) \\
\hline
$Q^{*2}$=4 & 2.213  & 0.6396 & 0.051(6)  &  2.72(13)   &14.04(71)  \\
           & 2.599  & 0.5884 & 0.098(2)  &  2.95(7)\0  &13.3(12)\0 \\
           & 2.972  & 0.5307 & 0.168(1)  &  1.90(13)   &10.51(17)  \\
           & 3.350  & 0.4554 & 0.281(3)  &  1.21(4)\0  &\07.61(41) \\
\hline\hline
\end{tabular}
\end{center}
\end{table}

\setcounter{table}{0}
\begin{table}[t]
\noindent \caption{Continued.}
\bigskip
\begin{center}
\begin{tabular}
{lcc|ccc} \hline\hline
$L^*$=0.505 & $T^*$ & $\rho^*$ & $D^*$ & $\eta^*$ & $\lambda^*$  \\
\hline\hline
$Q^{*2}$=0 & 1.638  & 0.5291 & 0.095(1) &2.02(6)\0  & \09.13(84)\\
           & 1.913  & 0.4891 & 0.141(2) &1.55(8)\0  & \07.62(42)\\
           & 2.190  & 0.4431 & 0.206(1) &1.10(6)\0  & \05.9(10)\0 \\
           & 2.476  & 0.3835 & 0.305(1) &0.79(4)\0  & \04.43(18)\\
\hline
$Q^{*2}$=1 & 1.652  & 0.5349 & 0.083(1) &2.29(14) & \09.97(74)\\
           & 1.924  & 0.4942 & 0.130(1) &1.67(6)\0  & \08.08(68)\\
           & 2.187  & 0.4474 & 0.193(1) &1.19(4)\0  & \06.48(18)\\
           & 2.509  & 0.3873 & 0.295(3) &0.81(5)\0  & \04.83(36)\\
\hline
$Q^{*2}$=2 & 1.728  & 0.5476 & 0.071(2) &2.80(10) & 10.8(11)\0  \\
           & 2.029  & 0.5049 & 0.118(1) &1.96(18) & \08.70(92)  \\
           & 2.288  & 0.4548 & 0.183(3) &1.34(13) & \07.04(71)  \\
           & 2.584  & 0.3926 & 0.283(1) &0.85(4)\0  & \04.93(40)\\
\hline
$Q^{*2}$=3 & 1.813  & 0.5643 & 0.058(1) &3.67(17) & 11.5(14)\0  \\
           & 2.102  & 0.5193 & 0.102(1) &2.40(13) & 10.6(11)\0  \\
           & 2.393  & 0.4691 & 0.166(2) &1.52(11) & \07.49(65) \\
           & 2.692  & 0.4041 & 0.266(4) &0.95(10) & \05.52(34) \\
\hline
$Q^{*2}$=4 & 1.922  & 0.5803 & 0.043(1) &5.28(74) & 14.94(28)   \\
           & 2.252  & 0.5335 & 0.092(2) &2.66(13) & 11.39(60)   \\
           & 2.541  & 0.4809 & 0.155(1) &1.71(10) & \08.52(58)  \\
           & 2.885  & 0.4135 & 0.259(3) &1.02(4)\0  & \05.80(24)\\

\hline\hline
\end{tabular}
\end{center}
\end{table}

\setcounter{table}{0}
\begin{table}[t]
\noindent \caption{Continued.}
\bigskip
\begin{center}
\begin{tabular}
{lcc|ccc} \hline\hline
$L^*$=0.6 & $T^*$ & $\rho^*$ & $D^*$ & $\eta^*$ & $\lambda^*$  \\
\hline\hline
$Q^{*2}$=0 & 1.475 &  0.4900 & 0.094(1)  & 1.76(8)\0 & \09.39(96) \\
           & 1.726 &  0.4521 & 0.140(1)  & 1.37(8)\0 & \07.67(37)\\
           & 1.948 &  0.4088 & 0.199(2)  & 1.02(6)\0 & \06.00(68)\\
           & 2.211 &  0.3520 & 0.294(3)  & 0.66(6)\0 & \04.13(45)\\
\hline
$Q^{*2}$=1 & 1.490 &  0.4947 & 0.084(1)  & 2.01(8)\0 & \08.92(39) \\
           & 1.731 &  0.4563 & 0.128(1)  & 1.45(8)\0 & \07.46(59) \\
           & 2.011 &  0.4116 & 0.193(2)  & 1.10(4)\0 & \06.09(37) \\
           & 2.233 &  0.3519 & 0.287(1)  & 0.70(3)\0 & \03.85(27) \\
\hline
$Q^{*2}$=2 & 1.552 &  0.5083 & 0.069(1)  & 2.42(11)   & 10.06(31) \\
           & 1.810 &  0.4682 & 0.112(1)  & 1.72(9)\0  & \08.07(74)  \\
           & 2.080 &  0.4214 & 0.177(2)  & 1.19(14)   & \06.09(35)  \\
           & 2.314 &  0.3622 & 0.268(1)  & 0.79(4)\0  & \03.6(11)\0 \\
\hline
$Q^{*2}$=3 & 1.610 &  0.5239 & 0.055(1)  & 3.04(26)   &  11.25(61)\\
           & 1.879 &  0.4819 & 0.096(1)  & 2.05(6)\0  & \09.40(67) \\
           & 2.137 &  0.4349 & 0.156(1)  & 1.38(9)\0  & \06.95(67) \\
           & 2.435 &  0.3758 & 0.250(3)  & 0.90(4)\0  & \04.97(38) \\
\hline
$Q^{*2}$=4 &1.725  &  0.5381 & 0.040(1)  & 4.84(23)   & 10.0(24) \\
           & 2.023 &  0.4944 & 0.087(1)  & 2.37(6)\0  & 10.23(71) \\
           & 2.284 &  0.4452 & 0.147(2)  & 1.52(5)\0  & \07.85(70)  \\
           & 2.608 &  0.3812 & 0.248(1)  & 0.99(7)\0  & \05.72(25)  \\
\hline\hline
\end{tabular}
\end{center}
\end{table}

\setcounter{table}{0}
\begin{table}[t]
\noindent \caption{Continued.}
\bigskip
\begin{center}
\begin{tabular}
{lcc|ccc} \hline\hline
$L^*$=0.8 & $T^*$ & $\rho^*$ & $D^*$ & $\eta^*$ & $\lambda^*$  \\
\hline\hline
$Q^{*2}$=0 & 1.234 &  0.4302 & 0.085(1) & 1.51(5)\0 & \07.38(51)\\
           & 1.426 &  0.3956 & 0.126(1) & 1.13(4)\0 & \06.21(28)\\
           & 1.632 &  0.3568 & 0.181(1) & 0.85(2)\0 & \04.93(13)\\
           & 1.856 &  0.3051 & 0.267(3) & 0.58(5)\0 & \03.64(23)\\
\hline
$Q^{*2}$=1 & 1.246 &  0.4364 & 0.074(1) & 1.69(1)\0 & \08.09(50)\\
           & 1.458 &  0.4016 & 0.115(1) & 1.23(7)\0 & \06.86(51)\\
           & 1.686 &  0.3602 & 0.173(1) & 0.89(8)\0 & \05.56(41)\\
           & 1.885 &  0.3059 & 0.259(1) & 0.60(4)\0 & \03.64(30)\\
\hline
$Q^{*2}$=2 & 1.308 &  0.4513 & 0.060(1) & 2.23(6)\0 & \09.52(13)\\
           & 1.514 &  0.4143 & 0.099(2) & 1.46(8)\0 & \07.22(22)\\
           & 1.712 &  0.3734 & 0.150(2) & 1.02(3)\0 & \05.74(9)\0\\
           & 1.933 &  0.3207 & 0.233(1) & 0.70(2)\0 & \04.04(24)\\
\hline
$Q^{*2}$=3 & 1.352 &  0.4666 & 0.047(1) & 2.91(12)   & \09.02(12)\\
           & 1.579 &  0.4290 & 0.082(3) & 1.85(5)\0  & \07.42(29)\\
           & 1.801 &  0.3858 & 0.136(2) & 1.24(6)\0  & \06.19(40)\\
           & 2.045 &  0.3319 & 0.219(1) & 0.78(6)\0  & \04.42(35)\\
\hline
$Q^{*2}$=4 & 1.447 &  0.4800 & 0.032(1) & 4.92(47)   &  10.81(15)\\
           & 1.674 &  0.4416 & 0.073(1) & 2.18(7)\0  & \08.87(40)\\
           & 1.914 &  0.3975 & 0.125(2) & 1.42(1)\0  & \07.22(39)\\
           & 2.187 &  0.3402 & 0.212(1) & 0.90(5)\0  & \04.65(41)\\
\hline\hline
\end{tabular}
\end{center}
\end{table}

\clearpage
\listoffigures
\clearpage

\end{document}